\documentclass[11pt]{article}
\usepackage{moriond,epsfig}
\def\Journal#1#2#3#4{{#1} {\bf #2}, #3 (#4)}
\def\mJournal#1#2#3#4#5{{#1}, #2 {\bf #3}, #4 (#5)}
\def\NIM{\em Nucl. Instrum. Methods}

\def\NPB{{\em Nucl. Phys.} B}

\def\PRL{\em Phys. Rev. Lett.}
\def\PRD{{\em Phys. Rev.} D}

\def\be{\begin{equation}}
\def\ee{\end{equation}}
\def\bea{\begin{eqnarray}}
\def\eea{\end{eqnarray}}

\def\DZERO{{\rm{DO}\hspace{-0.65em}/}\hspace{0.5em}}
\def\MET{{E_{T}\hspace{-1.0em}/}\hspace{0.5em}}
\begin{document}
\vspace*{4cm}
\title{SEARCH FOR HIGGS, LEPTOQUARKS, AND EXOTICS AT TEVATRON}

\author{ SONG MING WANG \footnote{ming@phys.ufl.edu} \\
(For the CDF and \DZERO Collaborations)}

\address{Department of Physics, University of Florida,\\
P.O. Box 118440, Gainesville, Florida 32611-8440, USA}

\maketitle\abstracts{This paper reviews some of the most recent results from the
CDF and \DZERO experiments on the searches for Standard Model and Non-Standard Model
Higgs bosons, and other new phenomena at the Tevatron.
Both experiments examine data from proton anti-proton collision
at $\sqrt{s} = 1.96$ TeV, of integrated luminosity $\sim 200$ pb$^{-1}$ (per experiment),
to search for Higgs predicted in the Standard Model and beyond Standard Model,
supersymmetric particles in the Gauge Mediated Symmetry Breaking scenario,
leptoquarks, and excited electrons. No signal was observed, and limits on
the signatures and models are derived.}

\section{Introduction}
The Standard Model (SM) has been remarkably confirmed by experiments over the
past decades. Its predictions have been tested to very high accuracy, and all the
particles predicted by the SM have been found, except the Higgs boson.
Given its success, however, there are hints that the SM is not a complete theory.
It cannot explain the mechanism behind the electroweak symmetry breaking,
no accounting for gravity, no prediction on the unification of the gauge couplings
at high energy scale, and recent cosmological observation indicates that the
SM particles only account for $\sim 4$\% of the matter of the universe.
Therefore extensions to the SM have been constructed to make the theory more
complete. Some of these extension models include Supersymmetry (SUSY),
Grand Unified Theory (GUT), and Extra Dimensions.
These extensions predict signs of new physics are very rare.
At CDF \cite{CDFdet} and \DZERO \cite{D0det} we search for these rare processes in
proton anti-proton collisions at $\sqrt{s}=1.96$ TeV, based on their
unique signatures that the processes display in its final states.
These final states are generally described in terms of physical signatures produced
by the particles involved. Example of these signatures from new physics are :
missing transverse energy ($\MET$), due to the presence of neutral and weakly interacting
particles such as neutrino or the lightest supersymmetric particle (LSP);
jets originating from the secondary vertices, due to the decay of Higgs, squarks,
or gluinos; photons , from the decay of the next lightest supersymmetric particle
(NLSP) in the Gauge Mediated Symmetry Breaking (GMSB) scenario.
In this paper we report on the recent results on Higgs, leptoquarks, and other
exotics searches performed by CDF and \DZERO based on $\sim 200$ pb$^{-1}$ data
(per experiment) collected since the start of Run 2 data taking in March 2001.

\section{Search for Higgs Boson}

\subsection{Search for Standard Model Higgs Boson in the Associated Production Channel}\label{subsec:WHlvbb}
\begin{figure}[htbp]
\includegraphics[width=0.47\textwidth]{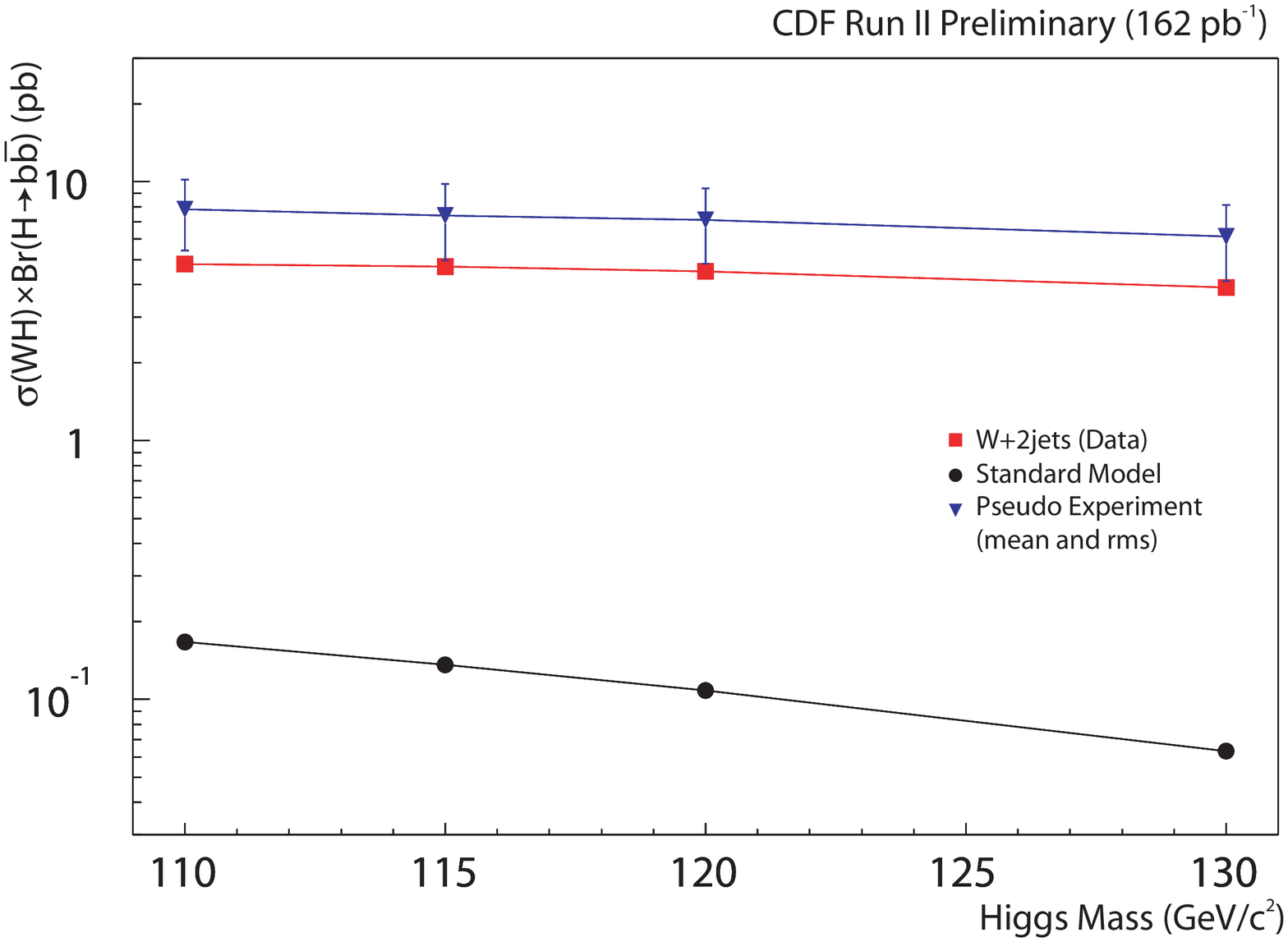}
\hfill
\includegraphics[width=0.47\textwidth]{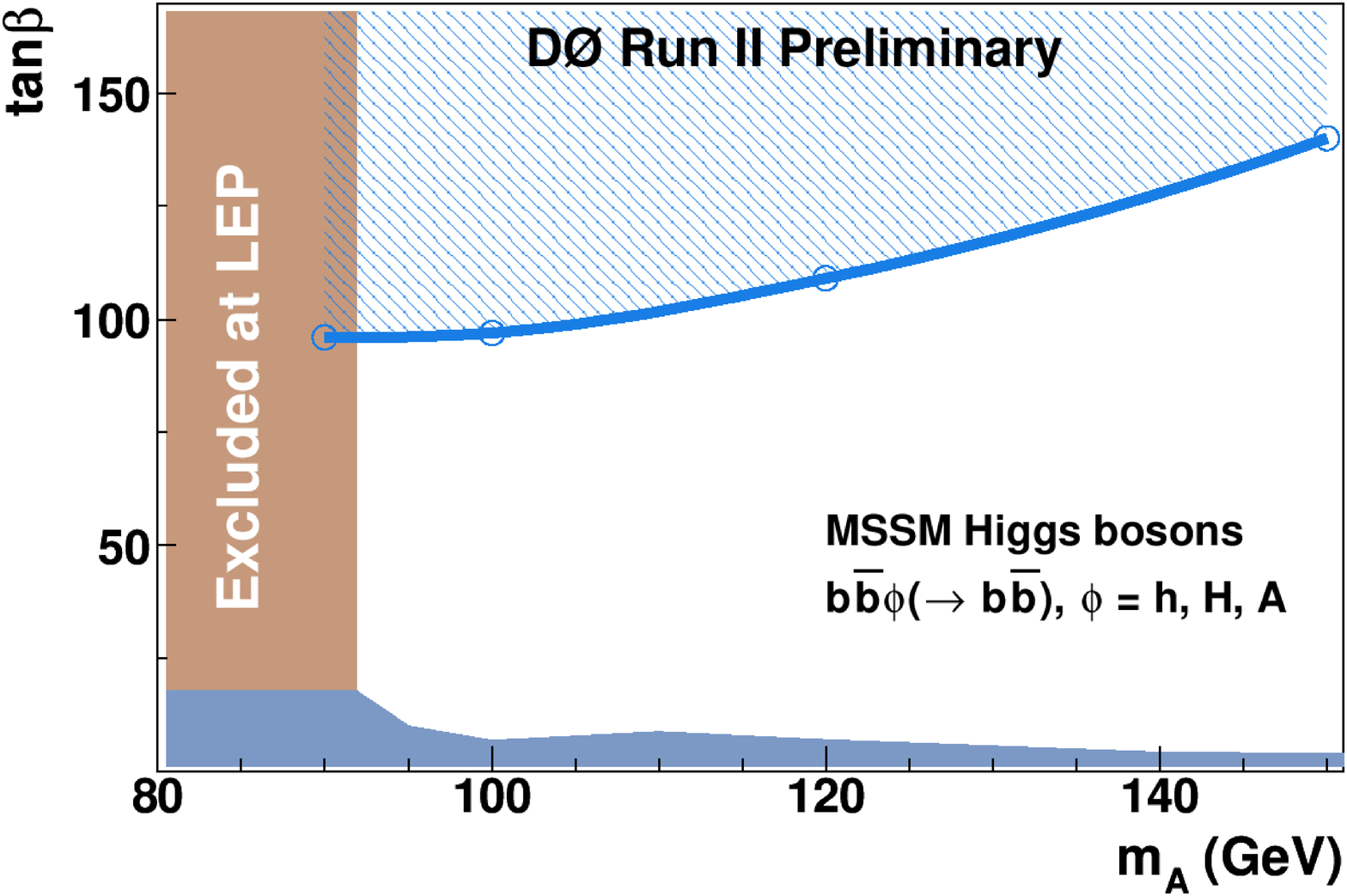}
\\
\parbox[t]{0.47\textwidth}{\caption{The 95\% C.L. upper limit (red squares) on the $W^{\pm}H$
cross section times branching ratio as a function of the Higgs boson mass.}
\label{fig:cdf_wh_limit}}
\hfill
\parbox[t]{0.47\textwidth}{\caption{The 95\% C.L. lower limit on $\tan{\beta}$ set as a function
of $\rm{m_{A}}$ (thick blue curve).}
\label{fig:d0_bbh_limit_2}}
\end{figure}

At Tevatron the dominant process for producing SM Higgs boson is through
gluon-gluon fusion ($gg \rightarrow H$) \cite{SMHiggsXS}. For light Higgs ($M_{H}<135$ GeV)
the most preferred decay channel is $H \rightarrow b\bar{b}$.
Unfortunately the search for Higgs boson in this production and decay channel
($gg \rightarrow H$, $H \rightarrow b\bar{b}$) will be swamped with
contribution from SM QCD multi-jet background.
So one of the channel that CDF has chosen to search for Higgs is the
$p\bar{p} \rightarrow W^{\pm}H \rightarrow l\nu b\bar{b}$ channel, where the
SM Higgs boson is produced in association with the $W$ boson. In the final state the
$W$ boson decays to a charged lepton and a neutrino, and the Higgs boson decays
to a $b\bar{b}$ pair.
These events are selected from a data sample of 162 pb$^{-1}$, by requiring
an energetic isolated electron (muon) in the central detector ($\mid \eta \mid < 1$)
with $E_{T} > 20$ GeV ($p_{T} > 20$ GeV), two jets of $E_{T} > 15$ GeV,
and large $\MET$ ($\MET > 20$ GeV).
At least one of these two jets is tagged as a b-jet. To suppress contribution from
$t\bar{t}$, events are removed if there are more than one charged lepton (electron or muon)
found in the event.
After applying all selection cuts, CDF observed 62 events, and the expected
number of SM background events is $60.55 \pm 4.43$. The majority of the background
contributions come from $Wb\bar{b}$, $Wc\bar{c}$, $Wc$, $t\bar{t}$, QCD, and events
with light flavored jets mis-tagged as b-jets. The 95\% C.L. upper limit on the
the cross section times branching ratio as a function of the Higgs mass is shown
in the Figure \ref{fig:cdf_wh_limit}. Currently no limit on the Higgs mass can be set as this
analysis is still limited by statistic.

\subsection{Search for MSSM Neutral Higgs Bosons at Large $\tan{\beta}$}\label{subsec:Hbbbb}
In the Minimum Supersymmetry extension of the Standard Model (MSSM) there are five
physical Higgs states after symmetry breaking. Of these five states, three states
are neutral. The coupling of these neutral Higgs bosons to the down-type quarks
are enhanced by a factor of $\tan{\beta}$ relative to the SM. Thus its production
rate is proportional to $\tan^{2}{\beta}$. The branching ratio of the neutral Higgs
to $b\bar{b}$ is expected to be large at large $\tan{\beta}$ ($\rm{BR \sim 90\%}$).

\DZERO has searched for the neutral Higgs bosons in the channel
$p\bar{p} \rightarrow Hb\bar{b} \rightarrow b\bar{b}b\bar{b}$ at high $\tan{\beta}$.
From a multi-jet data sample of 131 pb$^{-1}$, the analysis required an event to
have at least three b-tagged jets. Cuts of different transverse energy are applied
on the jets for searches at different Higgs mass points. The major SM contributions
in this analysis come from QCD multi-jet process, $t\bar{t}$, and 
$Z(\rightarrow b\bar{b}) + \rm{jets}$. With no evidence of neutral Higgs, the area
in the $\tan{\beta}$ vs $m_{A}$ ($m_{A}$ is the mass of the CP odd neutral Higgs boson)
space excluded by this search is shown in Figure \ref{fig:d0_bbh_limit_2}.

\subsection{Search for Non-SM Light Higgs Boson in $H \rightarrow \gamma\gamma$}\label{subsec:Hgg}
\begin{figure}[htbp]
\includegraphics[width=0.47\textwidth]{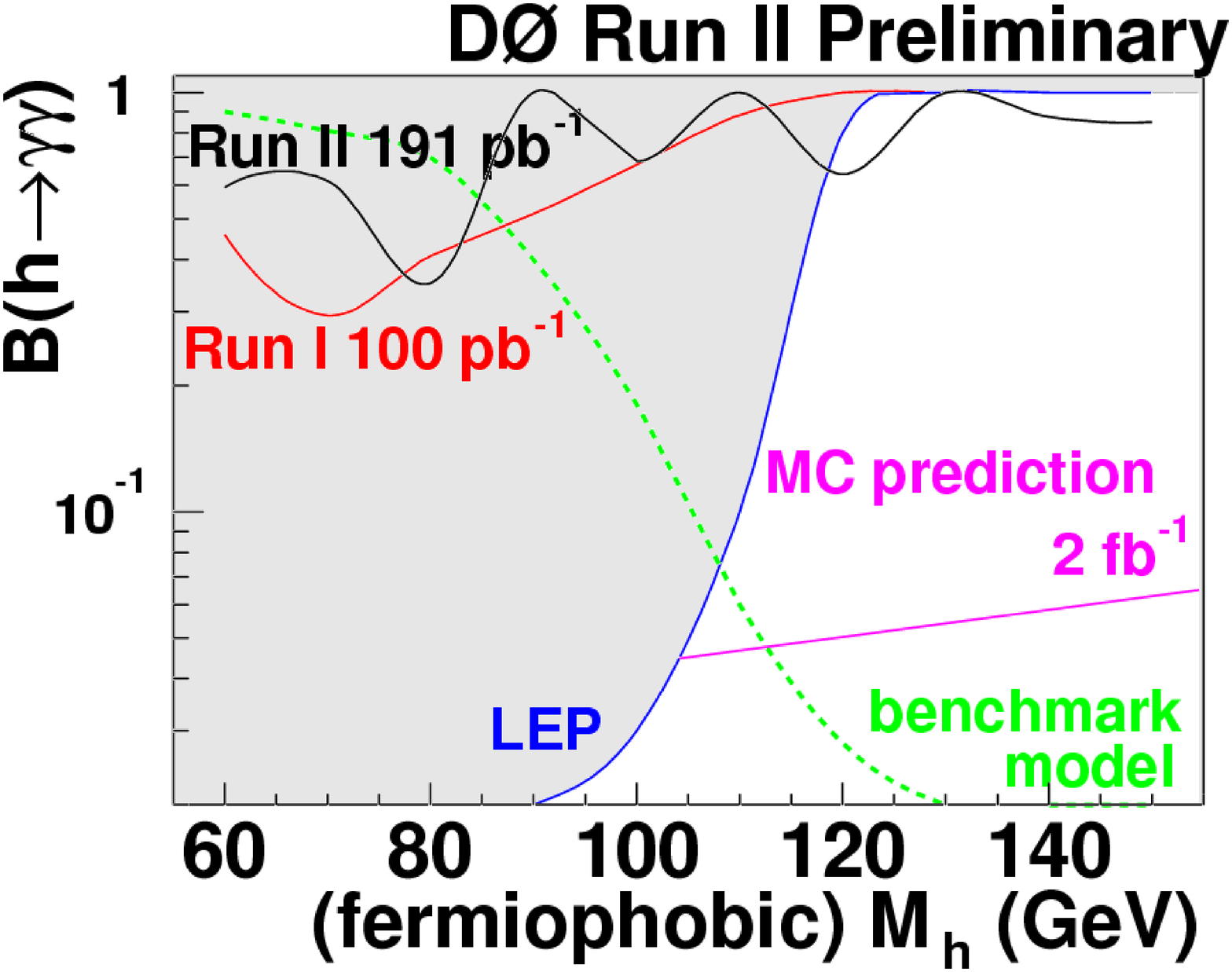}
\hfill
\includegraphics[width=0.47\textwidth]{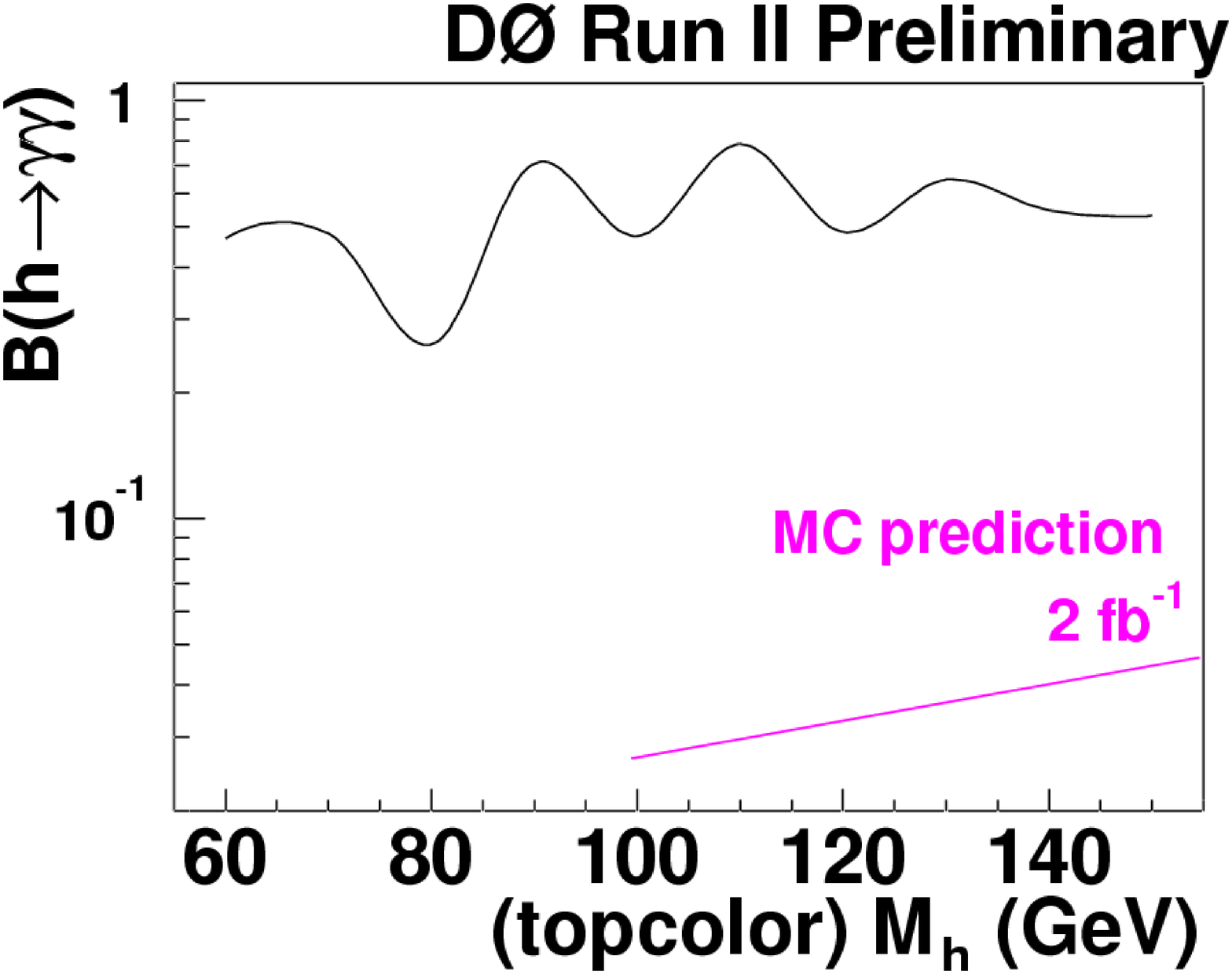}
\\
\parbox[t]{0.47\textwidth}{\caption{The 95\% C.L. limits on the Higgs decay branching fraction
 into two photons as function of Higgs mass for the Fermiophophic Higgs scenario.}
\label{fig:d0_hgg_BR_HF_2}}
\hfill
\parbox[t]{0.47\textwidth}{\caption{The 95\% C.L. limits on the Higgs decay branching fraction
 into two photons as function of Higgs mass for the Topcolor Higgs scenario.}
\label{fig:d0_hgg_BR_TC_2}}
\end{figure}

In some extension models to the SM, the coupling of the Higgs to the fermions are highly
suppressed and thus allow a large decay rate of $H \rightarrow \gamma\gamma$.
The search for Higgs in this decay channel $H \rightarrow \gamma\gamma$ has been
performed by the \DZERO experiment using Run 2 data sample of 191 pb$^{-1}$, under
two scenarios that enhance this decay channel. The first scenario is the Fermiophophic Higgs,
where the Higgs does not couple to fermions. The second scenario is the Topcolor Higgs,
where the top quark is the only fermion that the Higgs boson couples to.
In this analysis, events are selected with two electromagnetic clusters in the calorimeter,
with $E_{T}>25$ GeV, and pass photon identification. The vector sum transverse momentum
of the two clusters has to be greater than 35 GeV. The main SM background come from
QCD, where either jets are mis-identified as photons, and di-photon processes.
The dominant uncertainty in the background estimation comes from the measurement of the
photon mis-identification rate ($\sim 30\%$).
No evidence of signal is observed.
The 95\% C.L. limits on the Higgs decay branching fraction into two photons as function of
Higgs mass is shown in Figure \ref{fig:d0_hgg_BR_HF_2} for the Fermiophophic Higgs scenario,
and in Figure \ref{fig:d0_hgg_BR_TC_2} for the Topcolor Higgs scenario.

\subsection{Search for Doubly Charged Higgs Boson}\label{subsec:Hppmm}
\begin{figure}[htbp]
\includegraphics[width=0.47\textwidth]{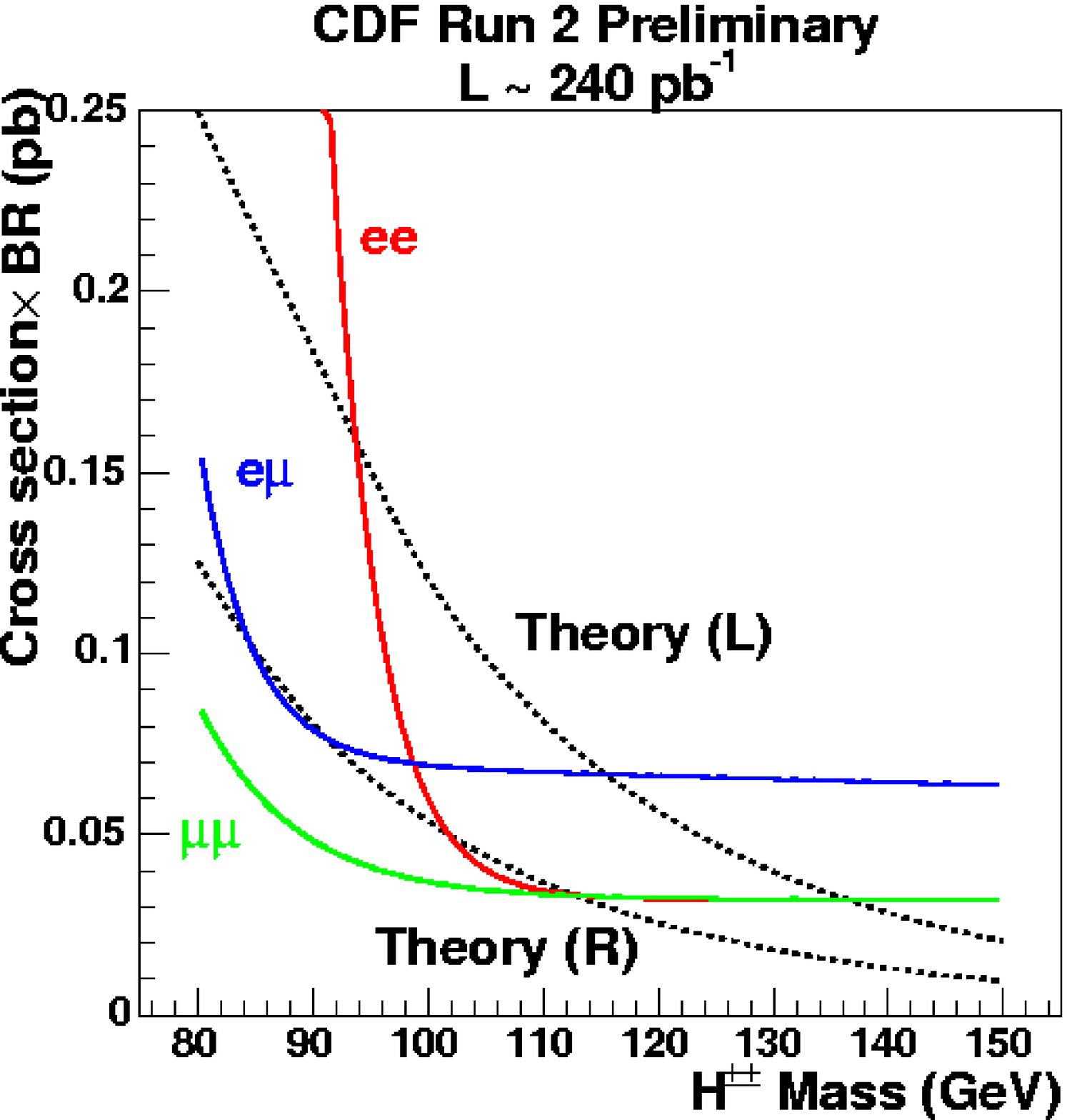}
\hfill
\includegraphics[width=0.47\textwidth]{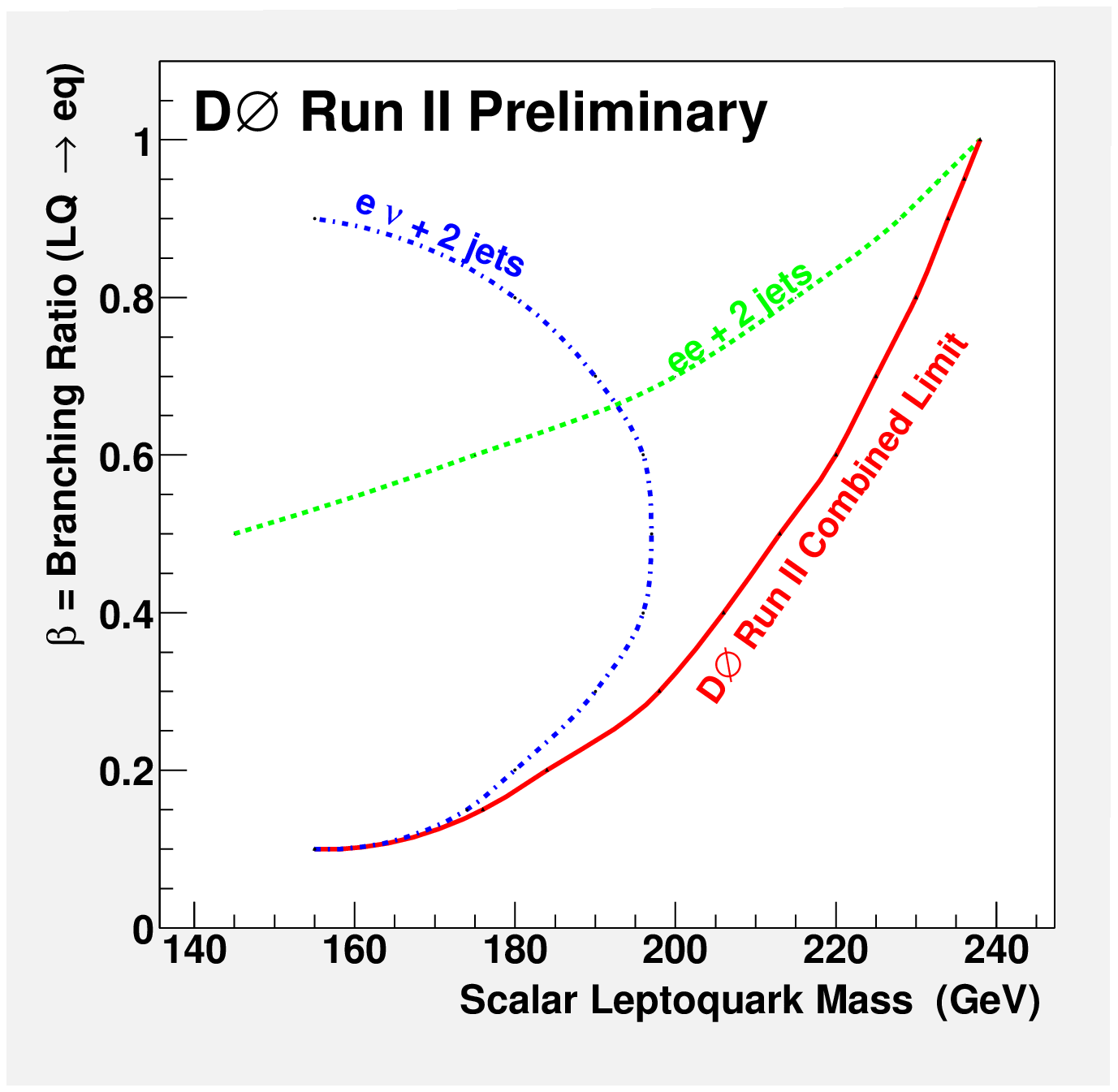}
\\
\parbox[t]{0.47\textwidth}{\caption{The 95\% C.L. cross section times branching ratio limit
 for pair-production of doubly-charged Higgs boson as function of the Higgs mass.}
\label{fig:cdf_Hpp_limit}}
\hfill
\parbox[t]{0.47\textwidth}{\caption{The 95\% C.L. lower limit on the mass of the first generation
 scalar leptoquark production as a function of $\beta$.}
\label{fig:d0_lq1_combine_limit_2b}}
\end{figure}

Doubly charged Higgs ($H^{\pm\pm}$) are predicted in models that contain Higgs triplets,
such as the Left-Right (L-R) symmetric models \cite{LRsymm}. In the SUSY L-R models \cite{SUSYLRsymm},
low mass $H^{\pm\pm}$ ($\sim 100$ GeV - 1 TeV) can exist in certain SUSY
phase space. In Run 2, CDF has searched for pair or single production of $H^{\pm\pm}$ 
at Tevatron, using a data sample of 240 pb$^{-1}$. This analysis searched for
 $H^{\pm\pm}$ decays to $H^{\pm\pm} \rightarrow e^{\pm}e^{\pm}$,
$H^{\pm\pm} \rightarrow \mu^{\pm}\mu^{\pm}$,
or $H^{\pm\pm} \rightarrow e^{\pm}\mu^{\pm}$.
The search is done in the mass window of $\pm 10\%$ of
a given Higgs mass ($\sim 3 \sigma$ of the detector resolution).
After the selection cuts, no event survived in the mass region $M_{H^{\pm\pm}} > 80$ GeV
($M_{H^{\pm\pm}} > 100$ GeV for electron pair channel), which is consistent with the
expectation from SM background prediction. The main SM contributions come from
jets mis-identified as charged leptons, $W+\rm{jets}$, $WZ$, and Drell-Yan.
Figure \ref{fig:cdf_Hpp_limit} shows the derived 95\% C.L. limit on the cross section
times branching ratio vs $M_{H^{\pm\pm}}$, for each same sign di-lepton channel.
\DZERO had performed similar analysis in the same sign di-muon final state,
with 106.6 pb$^{-1}$ of Run 2 data. The mass limit for the right-handed
(left-handed) doubly charged Higgs is $M_{H^{\pm\pm}} > 95$ GeV ($M_{H^{\pm\pm}} > 115$ GeV).

\section{Search for Leptoquarks}\label{sec:Leptoquarks}
Leptoquarks are color triplet bosons carrying both lepton and quark quantum numbers.
They are predicted in many extensions of SM (e.g. GUT, Technicolor, and
SUSY with R-parity violation) \cite{LQth1}. These particles can be scalar (spin=0) or vector (spin=1).
Leptoquarks may be produced in pairs in proton anti-proton collision, predominantly
through gluon fusion or quark antiquark annihilation. In the case of scalar leptoquark
pair production, the coupling of the leptoquarks to gluon is determined by the color
charge of the particles, thus it is model independent. However vector leptoquarks have
model independent trillinear and quadratic couplings to the gluon field.
The searches reported in this paper are only for scalar leptoquarks. These searches also
assume that leptoquarks only couple to lepton and quark of the same generation due to
experimental constraint of non observation of flavor changing neutral current process.
Leptoquark can decay either into a charged lepton and a quark ($\beta=1$), or into a
neutrino and a quark ($\beta=0$), where $\beta$ is the branching ratio of leptoquark decays into
a charged lepton and a quark. Therefore the final states in the pair production of
leptoquarks are $l^{\pm}l^{\mp}q\bar{q}$, $l^{\pm}\nu q\bar{q}$, and $\nu\bar{\nu}q\bar{q}$.

\subsection{First Generation Scalar Leptoquark : $e^{\pm}e^{\mp}q\bar{q}$ and $e^{\pm}\nu q\bar{q}$ channels.}
\label{subsec:LQ1ev}
\DZERO has searched for first generation scalar leptoquark in the channel
$LQ\bar{LQ} \rightarrow e^{\pm}e^{\mp}q\bar{q}$, and
$LQ\bar{LQ} \rightarrow e^{\pm}\nu q\bar{q}$, with data sample of 175 pb$^{-1}$.
The events with $e^{\pm}e^{\mp}q\bar{q}$ in the final state are selected by
requiring two electromagnetic clusters, with $E_{T}>25$ GeV, and at least two
jets, with $E_{T}>20$ GeV, reconstructed in the calorimeter.
The invariant mass from these two electron candidates should be outside the
Z boson mass window. The scaler sum of the transverse energy from the two electron
candidates and the two leading jets is required to be greater than 450 GeV,
to suppress contributions from QCD multi-jet, Z/Drell-Yan, and $t\bar{t}$ processes.
The events with $e^{\pm}\nu q\bar{q}$ in the final state are selected by
requiring an isolated electromagnetic cluster, with $E_{T}>35$ GeV and a track matched to the cluster,
and two jets with $E_{T}>25$ GeV in the event.
The selected event should have large $\MET$ ($\MET>30$ GeV).
The reconstructed transverse mass from the electron candidate and the neutrino should be
greater than 130 GeV, to suppress $W+\rm{jets}$ background. The scaler sum of the
transverse energy from the electron candidate, the $\MET$, and the two leading jets
is required to be larger than 330 GeV.
After applying all the selection cuts, the number of observed events and expected
SM background events from the $e^{\pm}e^{\mp}q\bar{q}$ ($e^{\pm}\nu q\bar{q}$) channel
is 0, and $0.4 \pm 0.1$ (2, and $4.7 \pm 0.9$), respectively.
The obtained scalar leptoquark mass limit at 95\% C.L. for the $e^{\pm}e^{\mp}q\bar{q}$
($e^{\pm}\nu q\bar{q}$) channel is $M>238$ GeV ($M>194$ GeV).
The results from these two channels are combined and the derived lower mass limit
as a function of $\beta$ is shown in Figure \ref{fig:d0_lq1_combine_limit_2b}.

CDF has also performed searches for pair production of first generation
scalar leptoquarks in these two channels. The 95\% C.L. mass limit are 230 GeV
for the $e^{\pm}e^{\mp}q\bar{q}$ channel, with 200 pb$^{-1}$ of data,
and 166 GeV for the $e^{\pm}\nu q\bar{q}$ channel, with 72 pb$^{-1}$ of data.

\begin{figure}[htbp]
\includegraphics[width=0.54\textwidth]{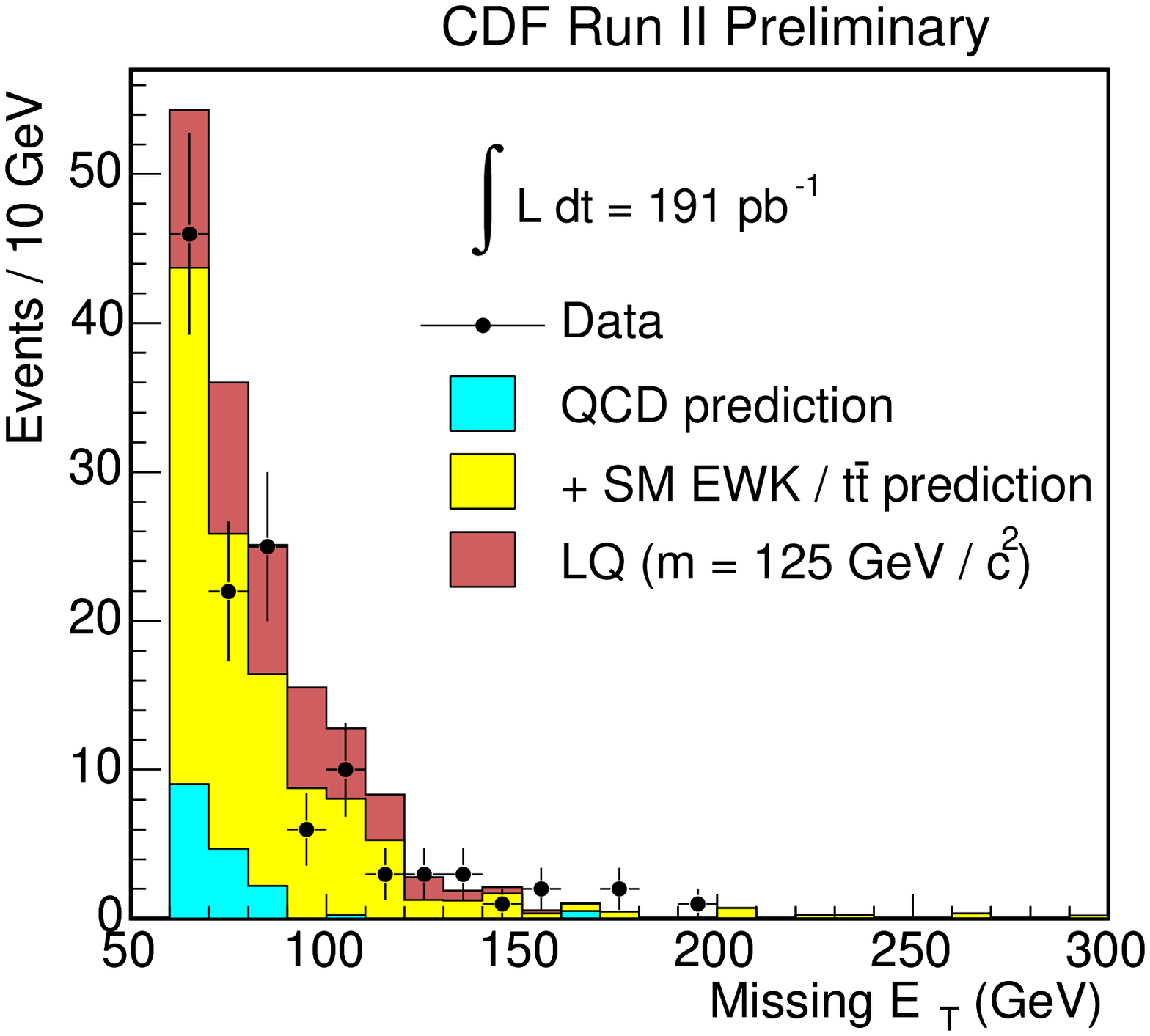}
\hfill
\includegraphics[width=0.47\textwidth]{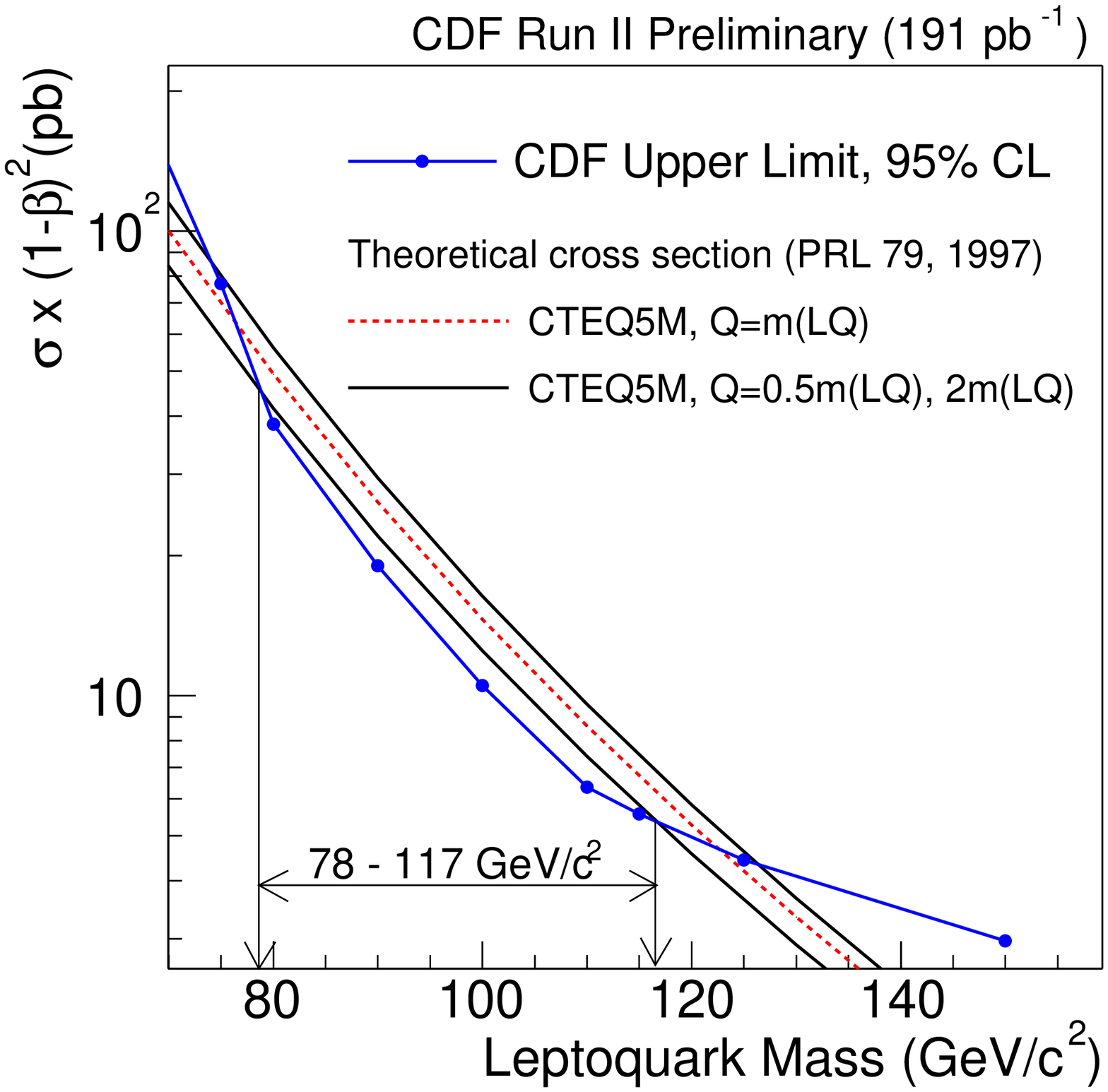}
\\
\parbox[t]{0.47\textwidth}{\caption{The $\MET$ distribution in the leptoquark
signal region (for the analysis $LQ\bar{LQ} \rightarrow \nu\bar{\nu}q\bar{q}$)
for data compared to SM background. Also shown is the expected distribution
arising from leptoquark production and decay at a mass of 125 GeV.}
\label{fig:cdf_lq_nunuqq_met}}
\hfill
\parbox[t]{0.47\textwidth}{\caption{The 95\% C.L. upper limit on the cross section times
squared branching ratio for scalar leptoquark pair production ($\beta = 0$).}
\label{fig:cdf_lq_nunuqq_limit}}
\end{figure}

\subsection{First Generation Scalar Leptoquark : $\nu\bar{\nu} q\bar{q}$ channel.}
\label{subsec:LQ1vv}
The pair production of first generation scalar leptoquarks has also been search
by CDF in the channel $LQ\bar{LQ} \rightarrow \nu\bar{\nu}q\bar{q}$, using a data
sample of 191 pb$^{-1}$. The events selected for this analysis contain two or three
jets and large $\MET$ ($\MET>60$ GeV) in the final state. The jets and $\MET$ directions
should not be aligned, to suppress QCD multi-jet events, which could fake large
$\MET$ due to jet energy mis-measurement. The events that have reconstructed charged
lepton (electron or muon) candidates are removed to reduce contributions from
$W/Z+\rm{jets}$ and $t\bar{t}$ processes. The number of events observed in the
signal region is 124, with an expected background of $118 \pm 14$ events.
Figure \ref{fig:cdf_lq_nunuqq_met} shows the $\MET$ distribution of the observed
events in the data, and the expected SM background contributions.
The 95\% C.L. limit on the cross section times squared branching ratio vs
the mass of leptoquark is shown in Figure \ref{fig:cdf_lq_nunuqq_limit}.
This analysis excludes the first generation scalar leptoquark mass in the range
between 78 GeV and 117 GeV. In Run 1 \DZERO \cite{D0LQrun1} had done a search in the same channel,
and the 95\% C.L. mass limit obtained is $M>98$ GeV.

\subsection{Second Generation Scalar Leptoquark}
\label{subsec:LQ2}
\begin{figure}[phtb]
\centerline{
    \resizebox{10cm}{!}{\includegraphics{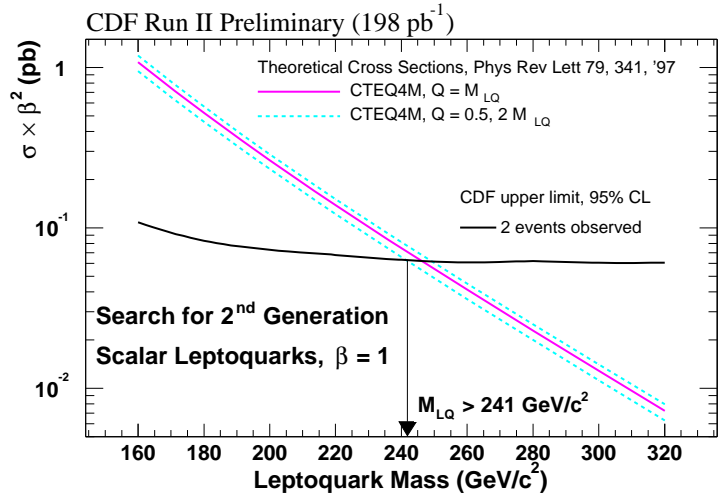}}
}
\caption{The {\tt MET45} trigger efficiency as a function of the corrected 
offline MET from the  {\tt JET50} sample.} 
\label{fig:cdf_lq2_limit}
\end{figure}

CDF has searched for second generation scalar leptoquarks decaying through the channel
$LQ\bar{LQ} \rightarrow \mu^{\pm}\mu^{\mp}q\bar{q}$. The integrated luminosity of the
data used in this analysis is 198 pb$^{-1}$. The events are selected by requiring the
presence of two opposite sign muon candidates with $p_{T}>25$ GeV, and two jets
(first leading jet $E_{T}>30$ GeV, and second leading jet $E_{T}>15$ GeV). The events
with reconstructed di-muon invariant mass below 15 GeV and between 76 GeV to 110 GeV
are vetoed to remove contributions from the decays of $J/\psi$, upsilon, and Z boson.
The contributions from Drell-Yan and $t\bar{t}$ processes are reduced with the cut on the
scaler sum of the transverse energy of the two leading jets, and the transverse momentum
of the two muon candidates. After applying all the selection cuts, two events survived
in the data sample, and the expected SM background is $3.2 \pm 1.2$.
The 95\% C.L. limit on the cross section times squared branching ratio vs
the mass of leptoquark is shown in Figure \ref{fig:cdf_lq2_limit}.
The leptoquark mass below 240 GeV is excluded by this analysis at 95\% C.L. .
\DZERO has also performed a similar search for second generation
scalar leptoquark in the same decay channel using Run 2 data of 104 pb$^{-1}$, and the
95\% C.L. mass limit obtained is $M>186$ GeV.

\section{Search for GMSB SUSY in $\gamma\gamma + \MET$}\label{sec:GMSB}
\begin{figure}[htbp]
\includegraphics[width=0.46\textwidth]{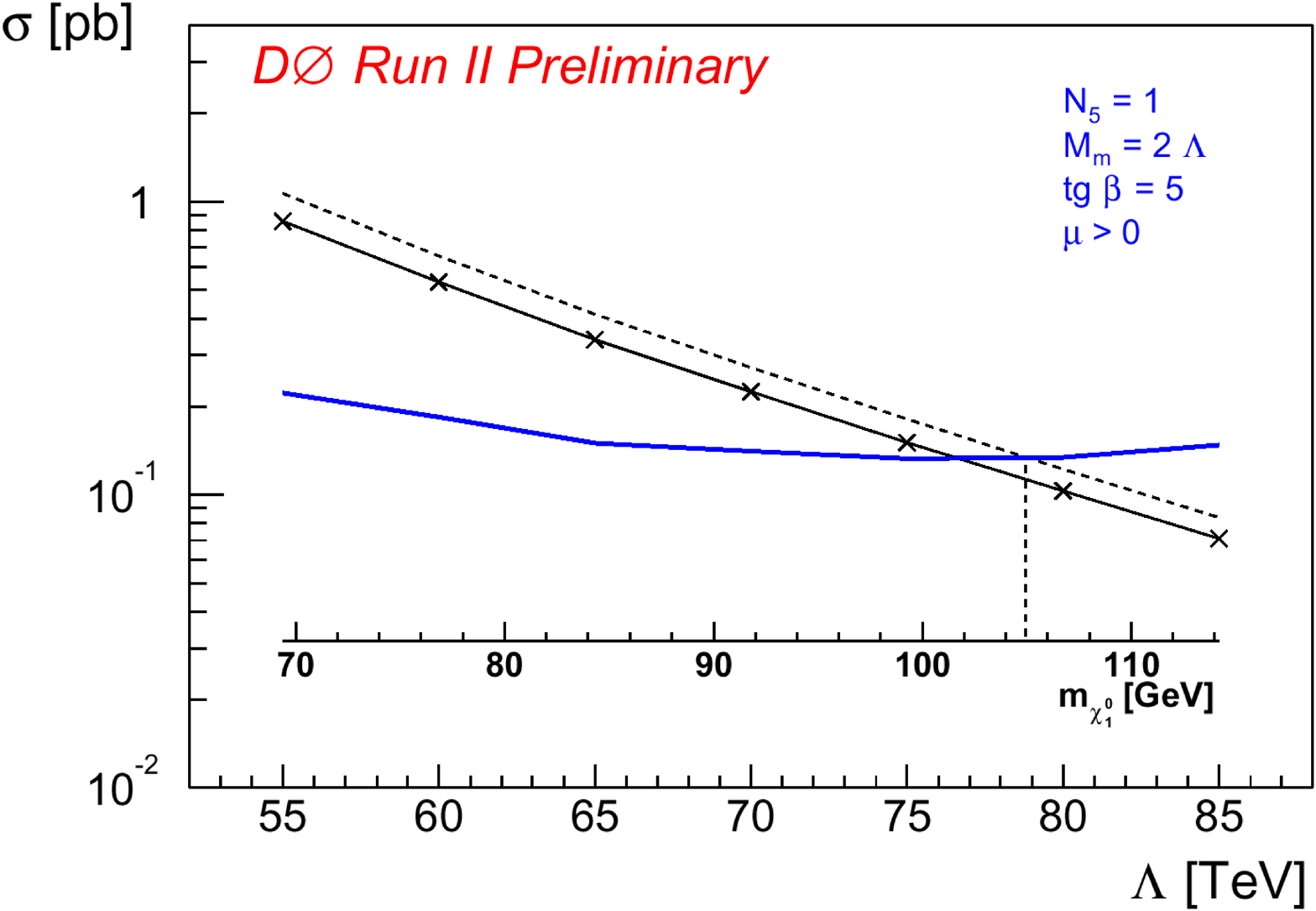}
\hfill
\includegraphics[width=0.46\textwidth]{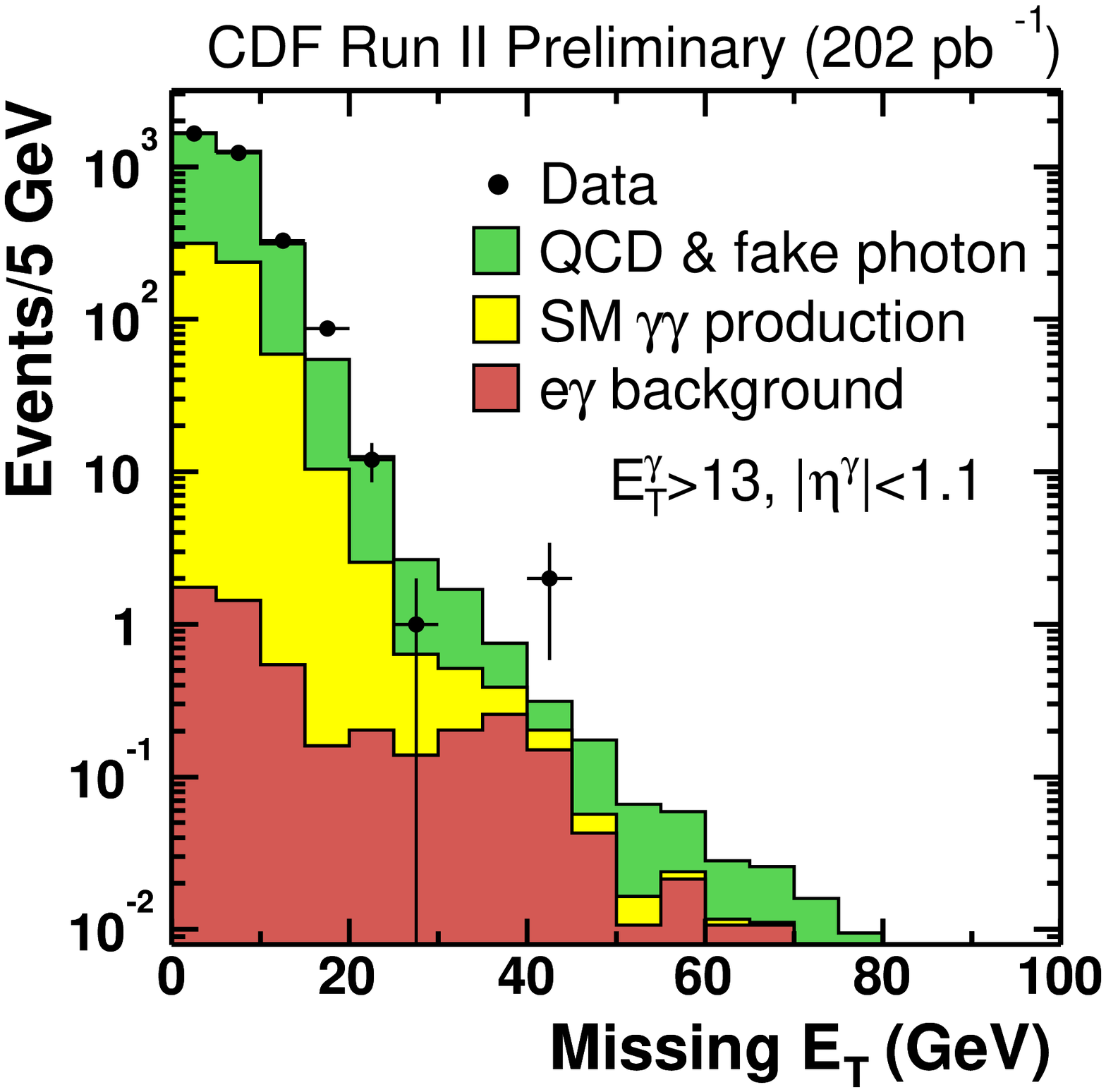}
\\
\parbox[t]{0.46\textwidth}{\caption{The 95\% C.L. upper limit on the GMSB cross section
 vs $\Lambda$. The solid black line is the LO theortical cross section, and the dashed
 black line is the LO theoretical cross section multiplied by $K$-factor.}
\label{fig:d0_gmsb_limit_2}}
\hfill
\parbox[t]{0.46\textwidth}{\caption{The $\MET$ distribution for di-photon events compared
 to the expected SM backgrounds for the GMSB SUSY search performed by CDF.}
\label{fig:cdf_gmsb_met_final_gg}}
\end{figure}

In the extended models with GMSB \cite{GMSB}, the lightest supersymmetric particle (LSP)
is the gravitino $\tilde{G}$. In the case where the neutralino $\tilde{\chi}^{0}_{1}$
is the next lightest symmetric particle (NLSP), it can decay to a $\tilde{G}$ by
emitting a photon ($\tilde{\chi}^{0}_{1} \rightarrow \gamma\tilde{G}$). 
The gravitino $\tilde{G}$ is neutral and interacts weakly, thus escapes detection.
If R-parity is conserved, this means that at least two $\tilde{\chi}^{0}_{1}$ will
be produced, and there will be at least two photons plus large $\MET$ in the final state.
The \DZERO analysis, based on Run 2 data sample of 185 pb$^{-1}$, required two
central photons ($\mid \eta \mid < 1.1$) with $E_{T} > 20$ GeV.
The largest SM background contribution comes from QCD with direct photon production,
or jets mis-identified as photons. From the study of optimizing the signal over
background, the $\MET > 40$ GeV region is chosen for limit calculation.
For this region 1 event is observed in the data, and the number of SM background
events is $2.5 \pm 0.5$. Using the values for the GMSB parameters of $M_{m}=2\Lambda$, $N_{5}=1$,
$\tan\beta=5$, and $\mu>0$, the 95\% C.L. upper limit on the GMSB cross section
vs $\Lambda$ is shown in Figure \ref{fig:d0_gmsb_limit_2}.
$\Lambda$ is the SUSY breaking scale parameter, $M_{m}$ is the messenger mass scale,
$N_{5}$ is the number of messenger fields, and $\mu$ is the Higgsino mass parameter.
The derived limit on $\Lambda$ is $\Lambda>78.8$ TeV, which corresponds to
$M_{\tilde{\chi}^{0}_{1}}>105$ GeV, and $M_{\tilde{\chi}^{\pm}_{1}}>191$ GeV.

Recent analysis from CDF, based on 202 pb$^{-1}$ data, set the 95\% C.L. limit
on $\Lambda$ to be $\Lambda>68$ TeV, $M_{\tilde{\chi}^{0}_{1}}>93$ GeV
and $M_{\tilde{\chi}^{\pm}_{1}}>168$ GeV.
The $\MET$ distribution of the selected events of this analysis is shown in
Figure \ref{fig:cdf_gmsb_met_final_gg}.

\section{Search for Excited Electron}\label{sec:ExEle}
\begin{figure}[htbp]
\includegraphics[width=0.47\textwidth]{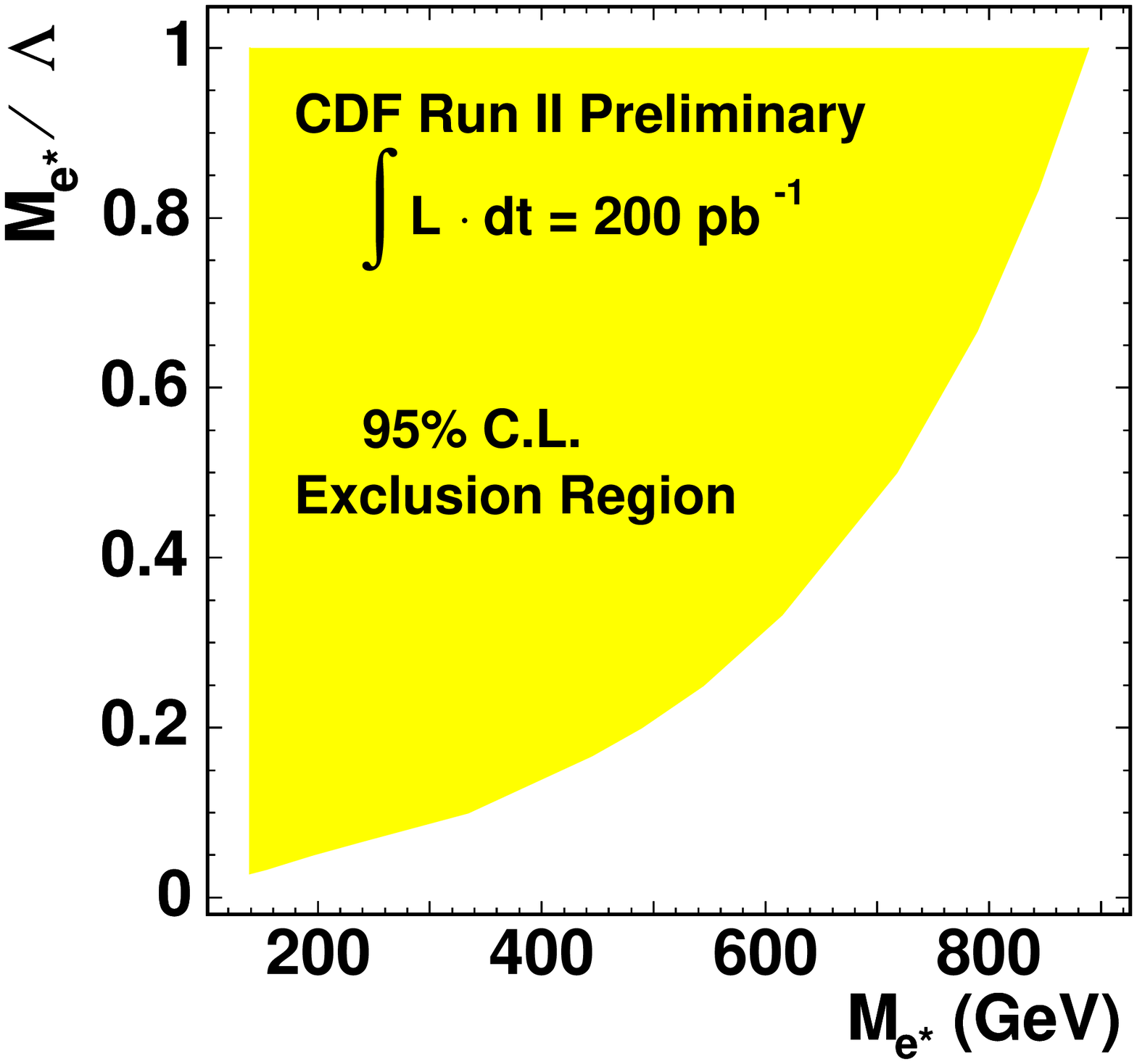}
\hfill
\includegraphics[width=0.47\textwidth]{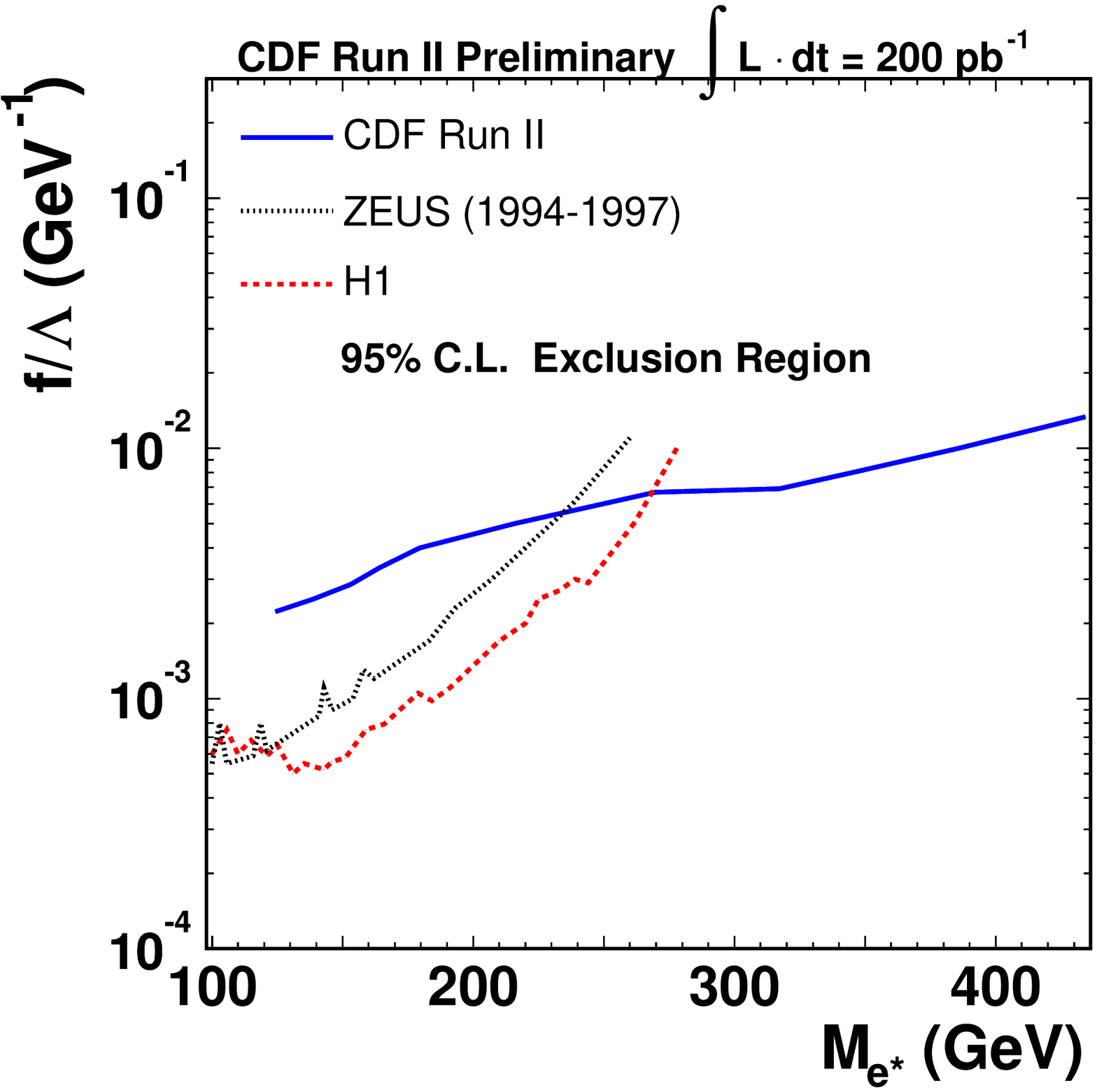}
\\
\parbox[t]{0.47\textwidth}{\caption{The 95\% C.L. exclusion region in $\rm{M_{e^{*}}}/\Lambda$
 vs $\rm{M_{e^{*}}}$ plane for contact interaction. $\rm{M_{e^{*}}}$ is the mass of the
excited electron, and $\Lambda$ is the compositeness scale.}
\label{fig:NinvmCI}}
\hfill
\parbox[t]{0.47\textwidth}{\caption{The 95\% C.L. exclusion region in $\rm{f}/\Lambda$
 vs $\rm{M_{e^{*}}}$ plane for gauge mediated interaction. $\Lambda$ is the compositeness scale,
 and f is the relative coupling strength to $\rm{SU(2)_{L}}$.}
\label{fig:GMfOverLambda}}
\end{figure}

The hierarchical structure of the fermion families in the SM may suggest that the
quarks and the leptons are composite particles that consist of more fundamental
entities. The observation of excited states of quarks and leptons will be a clear
sign that these particles are not elementary. At Tevatron excited electron ($e^{*}$)
could be produced through contact interactions or gauge mediated interactions.
The final state consists of an electron and an excited electron.
The excited electron then decays into an electron and a photon
($p\bar{p} \rightarrow ee^{*} \rightarrow ee\gamma$).
In Run 2 CDF searched for excited electron by selecting events with two electron
candidates and a photon candidate in the final state, from a data sample of 200 pb$^{-1}$.
The electron-photon invariant mass is reconstructed to check if there is a resonance.
The contributions from SM processes are mainly from $Z\gamma$+Drell-Yan, $Z+\rm{jets}$,
$WZ$, QCD multi-jet, and $\gamma\gamma + \rm{jets}$.
No excess of event is observed after applying all the selection cuts.
The derived 95\% C.L. exclusive regions for the cases of contact interactions and
gauge mediated interactions are shown in Figure \ref{fig:NinvmCI} and
\ref{fig:GMfOverLambda}, respectively.

\section{Summary}
The CDF and \DZERO Collaborations have performed searches for SM Higgs boson and physics
beyond the SM at Tevatron using the first 200 pb$^{-1}$ data collected since the
beginning of Run 2 data taking. No evidence of new physics has been observed so far,
and limits  on the signatures and models are derived. The new results are already as
competitive or better than Run 1 results.

\section*{References}

\end{document}